\documentclass[aps,prl,twocolumn]{revtex4}

\usepackage{graphicx}
\usepackage{dcolumn}
\usepackage{amsmath}
\usepackage{amssymb}
\usepackage{wasysym}
\usepackage{color}

\usepackage{mathtools}

\begin{document}

\title{Pseudogap in elemental plutonium}

\author{Mark~Wartenbe, Paul~H.~Tobash, John~Singleton, Laurel~E.~Winter, Scott Richmond and Neil~Harrison}

\affiliation{
Los~Alamos~National Laboratory, Los~Alamos,~NM~87545
}
\date{\today}













\section*{}
\noindent 

\begin{abstract}\noindent
Electronic correlations associated with incipient magnetism have long been recognized as an important factor in stabilizing the largest atomic volume $\delta$ phase of plutonium, yet their strength compared to those in the rare earths and neighboring actinides in the Periodic Table has largely remained a mystery. We show here using calorimetry measurements, together with prior detailed measurements of the phonon dispersion, that the $5f$ electrons of the $\delta$ phase reside in a pseudogapped state, accompanied by reductions in various physical properties below a characteristic temperature $T^\ast\approx$~100~K. The small characteristic energy scale of the pseudogapped state implies that the $5f$ electrons in plutonium are much closer to the threshold for localization and magnetic order than has been suggested by state-of-the-art electronic structure theory, revealing plutonium to be arguably the most strongly correlated of the elements.
\end{abstract}

\maketitle 



Plutonium (Pu) is located at an anomalously large discontinuity in volume between lighter and 
heavier elements in the actinide series of the Periodic Table, and itself undergoes significant discontinuous 
changes in volume with temperature $T$~\cite{moore2009}. Coulomb interactions have been shown to be an important factor in 
stabilizing the largest volume $\delta$ phase of Pu
relative to its other crystalline phases and those of neighboring elements~\cite{solovyev1991,soderlind1994,eriksson1999,bouchet2000,savrasov2001,soderlind2002,wills2004,sadigh2005,shick2005,shim2007,zhu2007,svane2007,soderlind2019}, yet these same interactions are expected to produce localized $5f$ electrons and consequent 
magnetism, as is often observed in rare-earth elements. 
While abundant evidence exists for fluctuating magnetic moments at $T\approx 1000$~K~\cite{migliori2016,janoschek2015,harrison2019}, 
at low $T$ it is as if Pu's magnetic moments are 
absent~\cite{lashley2005,mccall2006}. 
There is neither evidence for magnetic ordering, a phase transition, 
nor for free or screened magnetic moments producing large residual 
magnetic~\cite{harrison2019} or thermal~\cite{lawson2006} Gr\"{u}neisen parameters~\cite{kaiser1988}. 

\begin{figure}
\begin{center}
\includegraphics[width=0.95\linewidth]{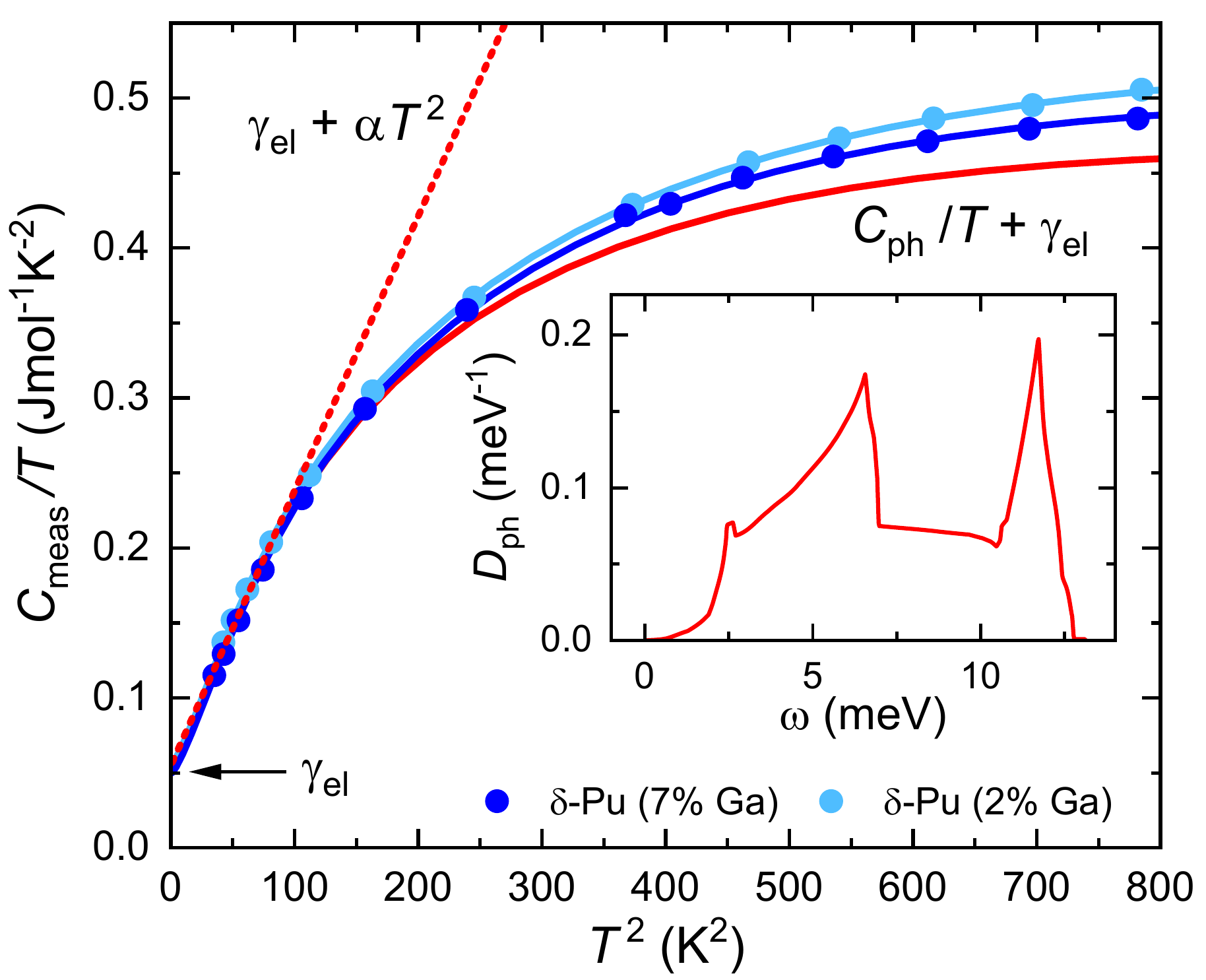}
\vspace{-4mm}
\textsf{\caption{
Measured heat capacity $C_{\rm meas}/T$ versus $T^2$ for $x=$~2\% and $x=$~7\% (data from this work;
light and dark blue circles).
The red dashed line is a fit of $(C_{\rm meas}/T)= (\gamma_{\rm el} + \alpha T^2)$
to the low-$T$ data.
The red solid curve is $(C_{\rm el}+C_{\rm ph})/T$, where
the phonon contribution to the heat capacity $C_{\rm ph}$ is calculated from the phonon density of states
shown in the inset~\cite{wong2003,mcqueeney2004} ($\omega$ is the phonon energy~\cite{supplemental}) and
the electronic part $C_{\rm el}$ is given by the conventional Sommerfeld
contribution~\cite{ashcroft1976} $\gamma_{\rm el}T$. Note that the curve undershoots the experimental data,
indicating that $\gamma_{\rm el} T$ is an underestimate of $C_{\rm el}$ at higher $T$.
}
\vspace{-9mm}
\label{phonon}}
\end{center}
\end{figure}

Contemporary proposals for the absence of magnetic moments in Pu include one, 
based on density functional theory (DFT)~\cite{soderlind2019}, in which the magnetic moments 
are disordered and exhibit a high degree of spin and orbital compensation, and another, based on dynamical mean field theory (DMFT)~\cite{savrasov2001,shim2007,zhu2007}, in which the moments are strongly hybridized with conduction electrons. 
It may be argued that either scenario can account for the fluctuating magnetic moments at an energy equivalent 
to $\approx$~1000~K reported in inelastic neutron scattering experiments~\cite{migliori2016,janoschek2015,janoschek2017,migliori2017}. 
However, both models significantly underestimate the unusually large Sommerfeld coefficient of 
$45 \lesssim\gamma_{\rm el}\lesssim 60$~mJmol$^{-1}$K$^{-2}$ of $\delta$-Pu~\cite{stewart1981,lashley2003,javorsky2006} 
(much larger than that of any other element) derived from calorimetry data. 
DFT and DMFT models of the electronic structure predict 
only modest values, 
$\gamma_{\rm el}\approx
7~{\rm mJmol^{-1}K^{-2}}$, 
for the Sommerfeld coefficient~\cite{shim2007,zhu2007,soderlind2019}. Hence, there is a need 
 to (i)~identify the fate of the missing magnetism in Pu, and (ii) explain why the relevant interactions have
 escaped being captured by the most advanced electronic structure theories. 

Here, we use low-noise calorimetry data combined with the results of prior high precision measurements of the phonon 
spectrum~\cite{wong2003,mcqueeney2004} to show that the magnetic moments in $\delta$-Pu, 
rather than being absent, are concealed 
below $T=T^\ast\approx$~100~K by their participation in 
a narrow pseudogapped state. The energy gap $\varepsilon_{\rm g}\approx$~12~meV separating peaks in the 
electronic density of states is found to be at least an order of magnitude lower than those predicted by DFT and DMFT~\cite{soderlind2019,savrasov2001,shim2007,zhu2007}. 
The hybridization between the $5f$ electrons of Pu and the conduction electrons is therefore much 
weaker than has generally been assumed; hence, the $5f$ electrons are close to the threshold for 
localization and Pu is arguably the {\it most strongly correlated of the elements}. 
The experimental signatures of the pseudogapped state are reminiscent of those in transition-metal 
oxides~\cite{timusk1999,lee2006,kim2014,kim2016,mott1976}, and include a peak in the electronic heat capacity 
below $T^\ast$, accompanied by a thermally activated Hall coefficient~\cite{brodsky1964} and a 
downturn below $T^\ast$ in physical quantities that are sensitive to the 
thermally averaged electronic density of states such as the electrical resistivity, Knight shift and magnetic susceptibility~\cite{joel1971,baclet2007,sandenaw1960,meotreymond1996,piskunov2009,piskunov2005}.

The 
$\delta$ phase of pure Pu normally exists at $T\gtrsim$~600~K, but can be stabilized down to
low $T$ by the substitution of small percentages of Am or group III elements 
such as Ga~\cite{hecker2004}; see Ref.~\cite{supplemental} for 
further details. 
Figure~\ref{phonon} shows an analysis of some of our heat-capacity $(C_{\rm meas})$
data for $\delta$-Pu$_{1-x}$Ga$_x$. 
At the lowest $T$, the data are fitted to~\cite{ashcroft1976} 
$(C_{\rm meas}/ T)= \gamma_{\rm el} + \alpha T^2,$
where $\alpha T^3$ is the leading order Debye phonon contribution to the heat capacity,
to obtain $\gamma_{\rm el}$ values in the range discussed above.
However,  the phonon contribution $C_{\rm ph}$
rapidly deviates from the Debye form as $T$ increases due to a combination of 
exceptionally soft acoustic-phonon branches and a high energy Einstein-like 
phonon mode in 
$\delta$-Pu~\cite{wong2003}.
Hence, $C_{\rm ph}$ is calculated from the phonon density of states previously determined 
from precision inelastic X-ray and neutron-scattering experiments~\cite{wong2003,mcqueeney2004,supplemental}.
The red curve in Fig.~\ref{phonon} shows $C = C_{\rm ph} +C_{\rm el}$
calculated in this way, assuming 
that the electronic contribution continues to behave as in conventional metals~\cite{ashcroft1976}, 
 {\it i.e.,} $C_{\rm el} =\gamma_{\rm el}T$.
 It is notable that the red curve {\it undershoots} the  
 data, suggesting that $C_{\rm el}$ grows 
 significantly larger than $\gamma_{\rm el}T$.

\begin{figure}
\begin{center}
\includegraphics[width=0.95\linewidth]{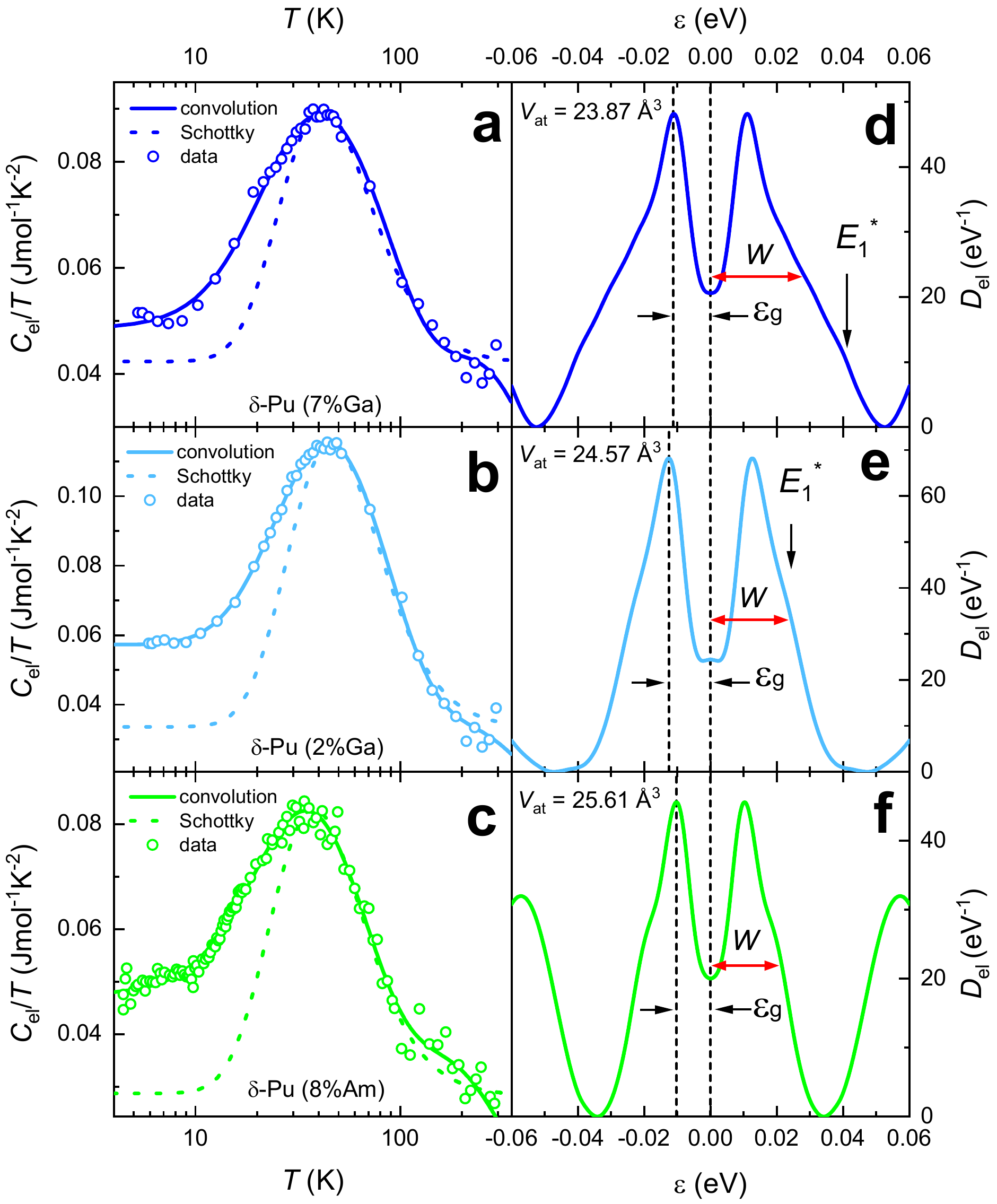}
\vspace{-3mm}
\textsf{\caption{
{\bf a}-{\bf c}, Electronic heat capacity divided by $T$, $C_{\rm el}/T$ (circles), 
versus $T$ for $x=$~7\% Ga, 2\% Ga and 8\% Am~\cite{javorsky2006}. 
Dotted lines are fits to the sum of a conventional Sommerfeld term and a Schottky term
(Eq.~6 of Re.~\cite{supplemental}); the reasonable fit at intermediate to high $T$
supports the
existence of a gapped electronic density of states $D_{\rm el}(\varepsilon)$. 
Solid curves are fits of Eq.~\ref{neilboris3} using
an iterative numerical procedure~\cite{supplemental} to obtain 
$D_{\rm el}(\varepsilon)$.
{\bf d}-{\bf f}, Plots of the  $D_{\rm el}(\varepsilon)$ obtained versus energy, $\varepsilon$, 
for $x=$~7\% Ga, 2\% Ga and 8\% Am, respectively, with the 
pseudogap energy $\varepsilon_{\rm g}$ and $W$, half the electronic subband width 
at half maximum, indicated. $V_{\rm at}$ refers to the approximate atomic 
volume at low $T$ for each composition. 
The increase in $D_{\rm el}(\varepsilon)$ at $|\varepsilon|\gtrsim2W$ likely 
originates from additional contributions to $C/T$ above $T \approx 150$~K, such as 
that from anharmonic phonons and the higher energy Schottky anomaly associated with the
invar effect~\cite{lawson2006}. Also shown in {\bf d}, {\bf e}  
is the activation energy $E_1^\ast$ 
from magnetostriction~\cite{harrison2019}.
}
\vspace{-8mm}
\label{bands}}
\end{center}
\end{figure}

In order to examine this atypically large electronic contribution,
$C_{\rm ph}$ deduced 
from inelastic X-ray and neutron scattering experiments~\cite{wong2003,mcqueeney2004,supplemental}
is subtracted from $C_{\rm meas}$; the results of this procedure
are shown in Fig.~\ref{bands}a-c.
As $T$ increases, $C_{\rm el}/T$ reaches a peak --- 
the thermodynamic hallmark of quasiparticles being thermally excited 
across a gap in the density of states~\cite{tari2003}.

Studies of $d$-electron 
systems~\cite{timusk1999,lee2006,kim2014,kim2016}
have shown
that when a pseudogap forms in a narrow-band system, 
a significant fraction of the electronic states that would normally reside 
at the Fermi energy pile up on either side of the gap, leading 
to peaks in the electronic density of states. 
Thermal excitations across a gap~\cite{tari2003} or pseudogap~\cite{timusk1999,lee2006,kim2014,kim2016}  
always lead to a peak in 
$C_{\rm el}/T$ at $T=T_{\rm peak}$, whose 
corresponding energy $k_{\rm B}T_{\rm peak}$ is a small fraction ($\approx0.3\times\varepsilon_{\rm g}$) 
of the gap energy $\varepsilon_{\rm g}$.
Therefore, the very similar behavior of $C_{\rm el}/T$ in Pu 
may well be because of the presence of a pseudogapped state. 
Under this assumption, the chemical potential is pinned to the middle
of the gap 
such that~\cite{supplemental};
\begin{equation}
\frac{C_{\rm el}}{T}=
\frac{R}{T}\int_{-\infty}^{\infty} D_{\rm el}(\varepsilon) \frac{\varepsilon^2}{k_{\rm B}^2T^2 }
\frac{{\rm e}^{\varepsilon/k_{\rm B}T}}{({\rm e}^{\varepsilon/k_{\rm B}T}+1)^2}{\rm d}\varepsilon.
\label{neilboris3}
\end{equation}
Here $\varepsilon$ is the quasiparticle energy, $R$ is the molar gas constant
and $D_{\rm el}(\varepsilon)$ is the quasiparticle density of states per Pu atom.
The fact that Eq.~\ref{neilboris3} assumes such a simple form ---
it is in effect a convolution of the heat capacity of a Schottky
anomaly\cite{tari2003} with $D_{\rm el}(\varepsilon)$ ---
allows an iterative numerical routine to be used to extract
$D_{\rm el}(\varepsilon)$ by fitting the experimental values of $C_{\rm el}/T$~\cite{supplemental}.
The resulting densities of states are shown in Fig.~\ref{bands}d-f.

We have confidence in this fitting
procedure because of the qualitative similarity of the peaks in $C_{\rm el}/T$ 
to the heat capacity of a Schottky anomaly~\cite{tari2003} (see dotted curves in
Figs.~\ref{bands}a-c and Ref.~\cite{supplemental}).
This implies that the widths $W$ of the peaks in 
$D_{\rm el}(\varepsilon)$ (Figs.~\ref{bands}d-f)
are sufficiently narrow so as not to obscure the gap, 
and further suggests that the gap remains robust over
$4 \lesssim T\lesssim 300$~K. 
 
\begin{figure}
\begin{center}
\includegraphics[width=1.00\linewidth]{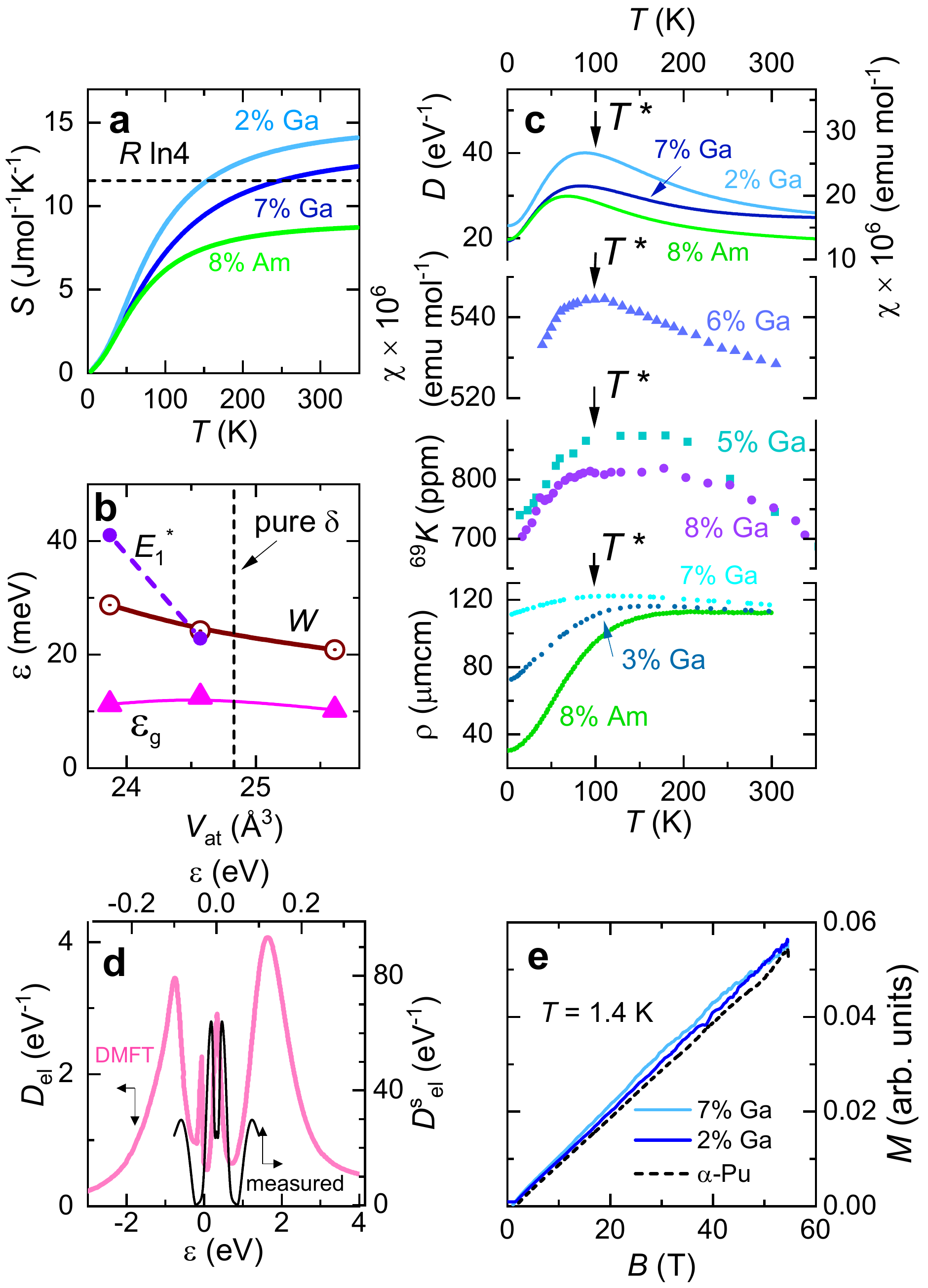}
\vspace{-3mm}
\caption{ {\bf a}, Calculated band entropy $S$ 
versus $T$ for each composition. {\bf b}, Electronic bandwidth $W$, pseudogap $\varepsilon_{\rm g}$ and 
previously determined\cite{harrison2019} excitation energy $E^\ast_1$ for each composition. 
{\bf c}, Calculated thermally-accessible electronic density of states $D(T)$ (Eq.~\ref{borisd})
for each composition (with the estimated spin susceptibility also shown, using $g=\frac{2}{7}$ and $J=\frac{1}{2}$, on the right-hand axis; see Methods), published susceptibility versus $T$\cite{meotreymond1996}, published NMR 
Knight shifts versus $T$\cite{piskunov2009,piskunov2005}, and published electrical resistivity\cite{joel1971,baclet2007} (divided by 3 for 8\% Am).
{\bf d},~Comparison of the $D_{\rm el}(\varepsilon)$ derived in this work
(black line and top and right-hand axes) for 2\% Ga against that calculated by way of dynamical mean field theory (DMFT) using quantum Monte Carlo (grey line and bottom and left-hand axes)\cite{zhu2007}. Arrows indicate the relevant axes. {\bf e},  Raw magnetization data for two Ga-stabilized $\delta-$Pu samples
and a pure $\alpha-$Pu sample in pulsed magnetic fields at $T=1.4$~K. For ease of comparison,
all samples were cut to have almost identical volumes~\cite{supplemental}. The similarity of the signals provides further evidence for a dominant Van~Vleck contribution~\cite{vanvleck1978,riseborough2000,supplemental}.}
\vspace{-8mm}
\label{pseudogap}
\end{center}
\end{figure}

Several attributes of the derived electronic density of states of $\delta$-Pu in 
Figs.~\ref{bands}d-f suggest the participation of $5f$-electrons. 
Firstly, the peaks in $D(\varepsilon)$ of each band are narrow, 
with the subbands in Figs.~\ref{bands}d-e having 
$W\approx$~20~meV. Such values are orders of magnitude 
lower than the $1 - 2$~eV normally associated with ordinary electronic bands, 
but are well within the range of values found in 
$d$- and $f$-electron systems~\cite{coleman2015}. 
Secondly, the total electronic entropy 
for each gapped band (Eq.~(7) in Ref.~\cite{supplemental}) is found
to saturate at a value close to $R\ln4$ (see Fig.~\ref{pseudogap}a). 
Such a large value occurs for a single electron per formula unit {\it only} when 
there are an equal number of thermally accessible filled and empty electronic states~\cite{tari2003};
this also supports our assumption that the chemical potential
does not vary much with $T$~\cite{supplemental}. 
Finally, the electronic bandwidths become progressively narrower when 
considered against the increasing atomic volume obtained by reducing the concentration of Ga and 
replacing it with Am~(Fig.~\ref{pseudogap}b). 
Such behavior is the expected result of a reduction in the transfer integrals of the 
$5f$ electrons hopping between adjacent atomic sites as the separation
of the Pu atoms increases. 
The absence of a proportionality of $W$ to the Ga concentration suggests that the presence of Ga and
any resulting disorder play a 
secondary role~\cite{supplemental}.
(We will return to this point in the discussion of Fig.~\ref{pseudogap}e below.)

Pseudogap signatures are not confined to the heat capacity~\cite{lee2006};  
in $d$-electron systems,
below a characteristic $T=T^\ast$, the pseudogap causes downturns in other physical properties, including 
the magnetic susceptibility $\chi$, nuclear-magnetic-resonance Knight shift 
and electrical resistivity $\rho$~\cite{lee2006,timusk1999}. 
Figure~\ref{pseudogap}c shows that this seems equally true for the same three
properties in $\delta$-Pu. 
Each is determined, at least in part, by 
the number of accessible quasiparticle
states at a particular $T$,
\begin{equation}
D(T) = \int_{-\infty}^{\infty}\frac{\partial f_{\rm D}(\varepsilon,T)}{\partial \varepsilon}D_{\rm el}(\varepsilon){\rm d} \varepsilon
\label{borisd}
\end{equation}
where $f_{\rm D}(\varepsilon,T)$ is the Fermi-Dirac distribution function
and $D_{\rm el}(\varepsilon)$ is the density of states already derived from the fits to the heat capacity.
The predictions of Eq.~\ref{borisd} are shown in Fig.~\ref{pseudogap}c; 
$D(T)$ consists of a broad maximum around $T^\ast \approx 100$~K
and a steep downturn for $T<T^\ast$. 
A close examination of $\chi$~\cite{meotreymond1996}, 
Knight shift (reflecting the local magnetic susceptibility at 
substituted $^{69}$Ga nuclei)~\cite{piskunov2009,piskunov2005}
and $\rho$~\cite{joel1971,baclet2007,sandenaw1960} data for $\delta-$Pu
(in Figs.~\ref{pseudogap}c) reveal that these too exhibit similar broad 
maxima around $T^\ast$ and steep downturns for $T<T^\ast$.
Therefore {\it all} are consistent with expected signatures of a pseudogap.

In the case of  $\chi$, the similarity 
in shape to $D(T)$ can be attributed to the proportionality of the contribution 
to $\chi$ from itinerant carriers to $D(T)$~\cite{ashcroft1976}; the large temperature-independent background, meanwhile, is consistent with significant ionic and band Van~Vleck contributions~\cite{vanvleck1978,riseborough2000,supplemental}. 
Qualitatively similar behavior occurs in the Knight shift (Fig.~\ref{pseudogap}c). 
 Here, consistency of the relaxation rate with free moment behavior $(T_1T)^{-1}\propto(T+T_\theta)^{1.5}$ at 
 $T>T^\ast$~\cite{piskunov2009} supports our proposal that the magnetic moments become 
 incorporated into a pseudogap state for $T<T^\ast$.
In the case of $\rho$, the similarity of its line shape to $D(T)$ suggests 
the possibility of there either being an approximate proportionality of the quasiparticle scattering 
rate to $D(T)$ ({\it i.e.,} Fermi's golden rule~\cite{pippard1989}) or a collapse in the 
Drude weight in response to a reduction~\cite{timusk1999} in $D(T)$ 
with decreasing $T$. Further evidence for a pseudogap in 
electrical transport may be provided by the thermally activated Hall effect, which indicates 
a 25-fold increase in the Hall coefficient~\cite{brodsky1964} at 
$T\lesssim T^\ast$~\cite{supplemental}.

Having determined the form of the electronic density of states 
of $\delta$-Pu (Figs.~\ref{bands}d-f) and verified that it is consistent with the $T-$dependence of
several physical properties (Fig.~\ref{pseudogap}c), it is useful to compare it with bandstructure 
measurements and calculations. Our finding of sharp spectral features 
at the Fermi energy is consistent with $T-$dependent 
photoemission experiments~\cite{joyce2011}.
However, the limited resolution $(35 \lesssim\Delta\varepsilon\lesssim 75$~meV) 
of photoemission measurements on $\delta$-Pu has thus far prohibited 
the observation of a pseudogap as small as $\approx 12$~meV, and is likely to 
have been a factor in the broader electronic bandwidth $2W\approx 200$~meV reported~\cite{joyce2011}.

When only photoemission experimental results are considered,
they strongly favor bandstructure calculations that assume 
the $5f$-electrons {\it a priori} to have an almost localized 
character~\cite{eriksson1999,wills2004,shim2007,zhu2007} 
over those that consider the $5f$ electrons to reside in broad electronic 
bands~\cite{solovyev1991,soderlind1994,bouchet2000,soderlind2002,soderlind2019}. 
In Fig.~\ref{pseudogap}d, we find that while the approximate forms of the 
pseudogap electronic densities of states in Figs.~\ref{bands}d-f resemble 
the results of DMFT calculations 
in which a strong hybridization between nearly localized $5f$ electrons and 
conduction electrons~\cite{zhu2007} produces a compensated Kondo 
semimetallic state~\cite{tutchton2020}, our experimentally determined bandwidths are roughly 
an order of magnitude narrower, and our densities-of-states are more than an order 
of magnitude higher. Indications are, therefore, that the degree of hybridization 
between the $5f$ electrons and the conduction electron states in 
$\delta$-Pu is considerably weaker than is generally 
considered in bandstructure calculations. 
Duality treatments in which $5f$ electrons are assumed to be partitioned 
between localized and itinerant states have shown that a weaker 
hybridization has the potential to produce sharper spectral features 
consistent with photoemission experiments~\cite{eriksson1999,wills2004}. 

A more localized $5f$ electron character 
implies that the 
magnetic properties are likely to be dependent on the lowest crystal electric 
field states~\cite{clementyev2009,supplemental}, and that interactions between magnetic 
moments on neighboring sites (as well as direct $f$-$f$ hopping) are more likely to be relevant in determining the 
pseudogap ground state in addition to the Kondo coupling between the magnetic moments 
and conduction electrons. 
Since geometric frustration is known to suppress antiferromagnetic 
long-range ordering in the fcc lattice~\cite{binder2980}, unconventional forms 
of magnetically correlated state may be present. 
The fcc lattice has, for example, been proposed to provide the ideal 
host for a three-dimensional variant of a resonant-valence-bond spin-liquid 
ground state~\cite{moessner2003}, thereby raising the possibility of a 
ground state with similarities to those considered in $d$-electron systems~\cite{lee2006}. 
The existence of a quantum-entangled state could potentially explain why extensive 
levels of frozen-in radiation damage are required to liberate free spins exhibiting 
Curie-Weiss behavior~\cite{mccall2006}, or why the pseudopgap remains 
robust under a magnetic field (Fig.~\ref{pseudogap}e). 
Observations consistent with a robust pseudogap state in a field include the 
non-magnetic ground state found in recent magnetostriction measurements extending 
to 15~T~\cite{harrison2019}, the small change of  $C_{\rm el}/T$ 
(and the corresponding $D_{\rm el}$ obtained by deconvolution~\cite{supplemental}) in a magnetic 
field of 14~T~\cite{lashley2005}, and the absence of any significant departures of the 
magnetization from linearity in fields of up to 55~T (Fig.~\ref{pseudogap}e~\cite{footnote}).

 The presence of strong electronic correlations only at temperatures $T<T^\ast$ 
 implies that the entropy (see Fig.~\ref{pseudogap}a) saturates quickly with increasing $T$, 
 giving rise to a large $T-$dependent contribution to the free energy $-TR\ln4$. 
 This is likely to enhance the stability of the $\delta$ phase relative to lower-volume phases at high $T$. 
 Our findings further suggest that the previously identified  $T-$scale of $\sim$~1000~K in 
 various experiments~\cite{harrison2019,lawson2006,janoschek2015} is unrelated to the low$-T$ 
 correlated electronic state. Rather, it relates to an electronic configuration that is accessed exclusively at high $T$~\cite{harrison2019}.

In summary, heat-capacity data with a high signal-to-noise ratio are used in conjunction
with an accurate determination of the phonon density of states~\cite{mcqueeney2004,wong2003}
to provide strong evidence for a pseudogap state in $\delta-$Pu.
This may be similar to those found in $d$-electron systems such as the cuprates 
and iridates~\cite{timusk1999,lee2006,kim2014,kim2016}; however, nothing analogous to this has 
been found previously in an element. 
Hence, the $5f$ electrons are much closer to the threshold of localization than has been 
suggested by electronic-structure models~\cite{solovyev1991,soderlind1994,eriksson1999,bouchet2000,savrasov2001,soderlind2002,wills2004,sadigh2005,shick2005,shim2007,zhu2007,svane2007,soderlind2019}, but this near localization 
is {\it not} associated with any form of conventional magnetic order.
Since the pseudogap is the only source of electronic entropy at low temperatures, it is 
solely responsible for the missing magnetic moments in $\delta$-Pu~\cite{lashley2005}. 
Of wider interest is whether the pseudogap in $\delta$-Pu is unique
or whether similar phenomena occur in Pu compounds that
exhibit anomalously high superconducting transition temperatures~\cite{sarrao2002}. 
Our findings warrant further studies by way of infrared optical spectroscopy or tunneling spectroscopy~\cite{riseborough2000}.

 \noindent
 {\bf Acknowledgments. }Support was provided by Los Alamos National Laboratory 
LDRD project 20180025DR. High magnetic field measurements  were supported by DoE
Basic Energy Science project "Science of 100 tesla." 
Some of this work was carried out at the National High Magnetic Field Laboratory, 
which is funded by NSF Cooperative Agreement 1164477, the State of Florida and DoE. 
We thank Mike Ramos for cutting the magnetization samples and Joe D. Thompson 
and Stephen Blundell for useful comments. MW acknowledges provision of a Seaborg Fellowship.

\bibliographystyle{naturemag}

\begin{thebibliography}{99}

\bibitem{moore2009} K. T. Moore, G. van de Laan, Nature of the 5$f$ states in actinide metals. {\it Rev. Mod. Phys.} {\bf 81}, 235-298 (2009).


\bibitem{solovyev1991} I. V. Solovye, A. I.  Liechtenstein, V. A. Gubanov, V. P. Antropov, O. K. Andersen, Spin-polarized relativistic linear-muffin-tin-orbital method - volume-dependent electronic-structure and magnetic-moment of plutonium. {\it Phys. Rev. B} {\bf 43}, 14414-14422 (1991).

\bibitem{soderlind1994} P. S\"{o}derlind, O. Eriksson, B. Johansson, J. M. Wills, Electronic-properties of $f$-electron metals using the generalized gradient approximation. {\it Phys. Rev. B} {\bf 50}, 7291-7294 (1994).


\bibitem{eriksson1999} O. Eriksson, J. N. Becker, A. V. Balatsky, J. M. Wills, Novel electronic configuration in $\delta$-Pu. {\it J. Alloy. \& Comp.} {\bf 287}, 1-5 (1999).

\bibitem{bouchet2000} J. Bouchet, B. Siberchicot, F. Jollet, A. Pasturel, Equilibrium properties of $\delta$-Pu: LDA`$+$~U calculations (LDA equivalent to local density approximation). {\it J. Phys.-Cond. Matter} {\bf 12}, 1723-1733 (2000).

\bibitem{savrasov2001} S. Y. Savrasov, G. Kotliar, E. Abrahams, Correlated electrons in $\delta$-plutonium within a dynamical mean-field picture. {\it Nature} {\bf 410}, 793-795 (2001). 

\bibitem{soderlind2002} P. S\"{o}derlind, A. Landa, B. Sadigh, Density-functional investigation of magnetism in $\delta$-Pu. {\it Phys. Rev. B } {\bf 66}, 205109 (2002).


\bibitem{wills2004} J. W. Wills, O., Eriksson, A. Delin, P. H. Andersson, J. J. Joyce, T. Durakiewicz, M. T. Butterfield, A. J. Arko, D. P. Moore, L. A. Morales, A novel electronic configuration of the $5f$ states in $\delta$-plutonium as revealed by the photo-electron spectra. {\it Journal of Electron Spectroscopy and Related Phenomena} {\bf 135}, 163-166  (2004).

\bibitem{sadigh2005} B. Sadigh, W. G. Wolfer, Gallium stabilization of $\delta$-Pu: Density-functional calculations. {\it Phys. Rev. B} {\bf 72}, 205122 (2005).

\bibitem{shick2005} A. B. Shick, V. Drchal, L. Havela, Coulomb-$U$ and magnetic-moment collapse in $\delta$-Pu. {\it Europhys. Lett.} {\bf 69}, 588-594 (2005).

\bibitem{shim2007} J. H. Shim, K. Haule, G. Kotliar, G., Fluctuating valence in a correlated solid and the anomalous properties of $\delta$-plutonium. {\it Nature} {\bf 446}, 513-516 (2007). 

\bibitem{zhu2007} J.-X. Zhu, A. K. McMahan, M. D. Jones, T. Durakiewicz, J. J. Joyce, J. M. Wills, R. C. Albers, Spectral properties of $\delta$-plutonium: Sensitivity to $5f$ occupancy, {\it Phys. Rev. B} {\bf 76}, 245118 (2007).

\bibitem{svane2007} A. Svane, L. Petit, Z. Szotek, W. M. Temmerman, Self-interaction-corrected local spin density theory of $5f$-electron localization in actinides. {\it Phys. Rev. B} {\bf 76}, 115116 (2007).

\bibitem{soderlind2019} P. S\"{o}derlind, A. Landa, B. Sadigh, Density-functional theory for plutonium. {\it Adv. Phys.} {\bf 68}, 1-47 (2019).


\bibitem{janoschek2015} M. Janoschek, P. Das, B. Chakrabarti, D. L. Abernathy, M. D. Lumsden, J. M. Lawrence, J. D. Thompson, G. H. Lander, J. N. Mitchell, S. Richmond, M. Ramos, F. Trouw, J.-X. Zhu, K. Haule, G. Kotliar, E. D. Bauer, 
The valence-fluctuating ground state of plutonium. {\it Sci. Adv.} {\bf 1}, e1500188 (2015); DOI: 10.1126/sciadv.1500188

\bibitem{migliori2016} A. Migliori, P. S\"{o}derlind, A. Landa, F. J. Freibert, B. Maiorov, B. J. Ramshaw, J. B. Betts, 
Origin of the multiple configurations that drive the response of $\delta$-plutonium's elastic moduli to temperature. {\it Proc. Nat. Acad. Sci. USA} {\bf 113}, 11158-11161 (2016). 

\bibitem{harrison2019} N. Harrison, J.~B. Betts, M.~R. Wartenbe, F.~F. Balakirev, S. Richmond , M. Jaime, P.~H. Tobash, Phase stabilization by electronic entropy in plutonium. {\it Nature Commun.} {\bf 10} 3159 (2019). 

\bibitem{lashley2005} J. C. Lashley, A. Lawson, R. J. McQueeney, G. H. Lander, Absence of magnetic moments in plutonium. {\it Phys. Rev. B} {\bf 72}, 054416 (2005). 

\bibitem{mccall2006} S. K. McCall, M. J. Fluss, B. W. Chung, M. W. McElfresh, D. D. Jackson, G. F. Chapline, Emergent magnetic moments produced by self-damage in plutonium. {\it Proc. Nat. Acad. Sci. USA} {\bf 103}, 17179-17183 (2006).

\bibitem{lawson2006} A. C. Lawson, J. A. Roberts, B. Martinez, M. Ramos, G. Kotliar, F. W. Trouw, M. R.  Fitzsimmons, M. P. Hehlen, J. C. Lashley, H. Ledbetter, R. J. Mcqueeney, A. Migliori, 
Invar model for $\delta$-phase Pu: thermal expansion, elastic and magnetic properties. {\it Phil. Mag.} {\bf 86}, 2713-2733 (2006). 

\bibitem{kaiser1988} A. B. Kaiser, P. Fulde, Giant magnetic Gr\"{u}neisen parameters in nearly ferromagnetic and heavy-fermion systems. {\it Phys. Rev. B} {\bf 37}, 5357-5363 (1988).

\bibitem{janoschek2017} M.~Janoschek, G.~Lander, J.~M.~Lawrence, J.~C.~Lashley, M.~Lumssden, D.~L., Abernathy, J.~D.~Thompson. Relevance of Kondo physics for the temperature dependence of the bulk modulus in plutonium. {\it Proc. Nat. Acad. Sci. USA} {\bf 114}, E268-E268 (2017). 

\bibitem{migliori2017} A. Migliori, P.~S\"{o}derlind, A. Landa, F.~J.~Freibert, B.~Maiorov, B.~J.~Ramshaw, J.~B.~Betts. The excited $\delta$-phase of plutonium. {\it Proc. Nat. Acad. Sci. USA} {\bf 114}, E269-E269 (2017). 


\bibitem{stewart1981} G. R. Stewart, R. O.  Elliott, {\it Conference Actinides
1981, Abstract Booklet} (Lawrence Berkeley Laboratory, Berkeley, CA, 1981), p. 206.

\bibitem{lashley2003} J. C. Lashley, J. Singleton, A. Migliori, J. B. Betts, R. A. Fisher, J. L. Smith, R. J. McQueeney, Experimental electronic heat capacities of $\alpha$- and $\delta$-plutonium: heavy fermion physics in an element. {\it Phys. Rev. Lett.} {\bf 91}, 205901 (2003). 

\bibitem{javorsky2006} P. Javorsk\'{y}, L. Havela, F. Wastin, E. Colineau, D. Bou\"{e}xi\`{e}re, heat capacity of $\delta$-Pu stabilized by Am. {\it Phys. Rev. Lett.} {\bf 96}, 156404 (2006).


\bibitem{wong2003} J. Wong, M. Krisch, D. F. Farber, F. Occelli, A. J. Schwartz, T.-C. Chiang, M. Wall, C. Boro, C., R. Xu, Phonon dispersions of fcc $\delta$-plutonium-gallium by inelastic X-ray scattering. {\it Science} {\bf 301}, 1078-1080 (2003).

\bibitem{mcqueeney2004} R. J. McQueeney,, A. C. Lawson, A. Migliori, T.M. Kelley, B. Fultz, M. Ramos,
B. Martinez, J. C. Lashley, S. Vogel, Unusual phonon softening in $\delta$-phase plutonium. {\it Phys. Rev. Lett.} {\bf 92}, 146401 (2004).


\bibitem{mott1976} N. F. Mott, Metal-insulator transition. {\it Rev. Mod. Phys.} {\bf 40}, 677-683 (1976).

\bibitem{timusk1999} T. Timusk, B. Statt, The pseudogap in high-temperature superconductors: an experimental survey. {\it Rep. Prog. Phys.} {\bf 62}, 61-122 (1999).

\bibitem{lee2006} P. A. Lee, N. Nagaosa, X.-G. Wen, Doping a Mott insulator: Physics of high-temperature superconductivity. {\it Rev. Mod. Phys.} {\bf 78}, 17-85 (2006).

\bibitem{kim2014} Y.~K Kim, O. Krupin, J. D. Denlinger, A. Bostwick, E. Rotenberg, Q. Zhao, J. F. Mitchell, J. W., Allen, B. J. Kim, Fermi arcs in a doped pseudospin-1/2 Heisenberg antiferromagnet. {\it Science} {\bf 345}, 187-190 (2014).

\bibitem{kim2016} Y. K. Kim, N. H. Sung, J. D. Denlinger, B. J. Kim, Observation of a $d$-wave gap in electron-doped Sr$_2$IrO$_4$. {\it Nature Physics} {\bf 12}, 37-42 (2016).

\bibitem{brodsky1964} M. B. Brodsky, Hall effect in plutonium. {\it Phys. Rev} {\bf 137}, A1423-A1428 (1964).


\bibitem{joel1971} J. Joel, C. Roux, M. Rapin, Resistivite electrique des solutions solides d'alliages Pu-Ga en phase $\delta$ a tres basses temperatures (4,2-300 $^\circ$K). {\it Journal of Nuclear Materials} {\bf 40}, 297-304 (1971). 

\bibitem{baclet2007} N. Baclet, M. Dormeval, L. Havela, J. M. Fournier, C. Valot, F. Wastin, T. Gouder, E. Colineau, C. T. Walker, S. Bremier, C. Apostolidis, G. H. Lander, Character of $5f$ states in the Pu-Am system from magnetic susceptibility, electrical resistivity, and photoelectron spectroscopy measurements. {\it Phys. Rev B} {\bf 75}, 035101 (2007).

\bibitem{sandenaw1960} T. A. Sandenaw, Heat capacity, thermal expansion and electrical resistivity of an 8~$a/o$ aluminum-plutonium (delta-phase stabilized) alloy below 300~$^\circ$K. {\it J. Phys. Chem. Solids} {\bf 16}, 329-336 (1960). 

\bibitem{meotreymond1996} S. M\'{e}ot-Reymond, J. M. Fournier, Localization of 5f electrons in plutonium: Evidence for the Kondo effect. {\it J. Alloys and Compounds} {\bf 232}, 119-125 (1996).

\bibitem{piskunov2005} Y. Piskunov, K. Mikhalev, A. Gerashenko, A. Pogudin, V. Ogloblichev, S. Verkhovskii, A. Tankeyev, V. Arkhipov, Yu. Zouev, S. Lekomtsev, Spin susceptibility of Ga-stabilized  $\delta$-Pu probed by 
$^{69}$ Ga NMR. {\it Phys. Rev. B} {\bf 71}, 174410 (2005).

\bibitem{piskunov2009} Y. Piskunov, K. Mikhalev, A. Buzlukov, A. Gerashenko, S. Verkhovskii,
V. Ogloblichev, V. Arkhipov, A. Korolev, Yu. Zouev, I. Svyatov, {\it Journal of Nuclear Materials} {\bf 385}, 25-27 (2009).


\bibitem{hecker2004} S. S. Hecker, D. R. Harbur, T. G. Zocco, Phase stability and phase transformations in Pu-Ga alloys. {\it Prog. Mater. Science} {\bf 49}, 429-485 (2004).

\bibitem{supplemental} See Supplemental Material at http://link.aps.org/supplemental/\dots. This document discusses the samples, heat capacity and magnetization measurements, the modeling of the phonon contribution, the extraction of the electronic density of states via deconvolution, thermodynamic functions of states such as entropy, tests of the deconvolution procedure, the effects of an asymmetric density of states, modeling of the Hall effect, a discussion of the $5f$ electron configurations and crystal electric fields, the effects of dopants and impurities, and the signal-to-noise limitations of prior measurements, and includes Refs. \cite{wolfer2000,moriya2009,fazekas1999,booth2012,sundermann2018,phelan2014,akintola2018}. 


\bibitem{wolfer2000} W. G. Wolfer, Radiation Effects in Plutonium: What is known? Where should we go from here? {\it Los Alamos Science} {\bf 26}, 274-285 (2000); https://lib-www.lanl.gov/lascience26.shtml

\bibitem{moriya2009} Y.~Moriya, H. Kawaji, T. Atake, M. Fukuhara, H. Kimura, A. Inoue, 
Heat capacity measurements on a thin ribbon sample of Zr$_{0.55}$Al$_{0.10}$Ni$_{0.05}$Cu$_{0.30}$ glassy alloy and Apiezon N high vacuum grease using a Quantum Design Physical Property Measurement System. {\it Cryogenics} {\bf 49}, 185-191 (2009).

\bibitem{fazekas1999} P.~Fazekas, {\it Lecture Notes on Electron Correlation and Magnetism} (World Scientific, 1999).

\bibitem{booth2012} C. H. Booth, Y. Jiang, D. L. Wang, J.N. Mitchell, P. H. Tobash, E. D. Bauer, M. A. Wall, P. G. Allen, D. Sokaras, D. Nordlund, T.-C. Weng, M. A. Torrez, and J. L. Sarrao, Multiconfigurational nature of $5f$ orbitals in uranium and plutonium intermetallics. {\it Proc. Natl. Acad. USA} {\bf 109}, 10205 (2012).

\bibitem{sundermann2018} M. Sundermann, H. Yavas, K. Chen, D. J. Kim, Z. Fisk, D. Kasinathan, M. W. Haverkort,
P. Thalmeier, A. Severing, and L. H. Tjeng, $4f$ crystal field ground state of the strongly correlated topological insulator SmB$_6$. {\it Phys. Rev. Lett.} {\bf 120}, 016402 (2018).

\bibitem{phelan2014} W. A. Phelan, S. M. Koohpayeh, P. Cottingham, J. W. Freeland, J. C. Leiner, C. L. Broholm, T. M. McQueen, {Correlation between bulk thermodynamic measurements
and the low-temperature-resistance plateau in SmB$_6$}. {\it Phys. Rev. X} {\bf 4}, 031012 (2014).

\bibitem{akintola2018} K. Akintola, S. R. Dunsiger, A. C. Y. Fang, M. Potma, S. R. Saha, X. Wang, J. Paglione, J. E. Sonier, Freezing out of a low-energy bulk spin exciton in SmB$_6$. {\it npg Quantum Materials} {\bf 3}, 36 (2018); doi:10.1038/s41535-018-0110-7





%
%
%






\bibitem{ashcroft1976} N. W. Ashcroft, N. D. Mermin,  {\it Solid state physics} (Saunders College Publishing, Orlando 1976).






\bibitem{tari2003} A.~Tari, The heat capacity of matter at low temperatures (Imperial College Press, 1$^{\rm st}$ ed., 2003).

\bibitem{coleman2015} P. Coleman, Introduction to many-body physics (Cambridge University Press, Cambridge 2015).

\bibitem{vanvleck1978} J. H. Van~Vleck, Quantum mechanics: The key to understanding magnetism. {\it Science} {\bf 201}, 113-120 (1978).

\bibitem{riseborough2000} P. S. Riseborough, Heavy fermion semiconductors. {\it Adv. Phys.} {\bf 49}, 257-320 (2000).


\bibitem{pippard1989} A.B. Pippard, Magnetoresistance in Metals (Cambridge University Press, Cambridge, UK, 1989).

\bibitem{joyce2011} J. J. Joyce,  T. Durakiewicz, K. S. Graham, E. D. Bauer, D. P. Moore, J. N. 
Mitchell, J. A. Kennison, R. L. Martin, L. E. Roy, G. E. Scuseria, Pu electronic structure and photoelectron spectroscopy. {\it Journal of Physics: Conference Series} {\bf 273}, 012023 (2011). 






\bibitem{tutchton2020} R. M. Tutchton, W. Chiu, R. C. Albers, G. Kotliar, J.-X. Zhu, Electronic correlation induced expansion of Fermi pockets in $\delta$-plutonium. {\it Phys. Rev. B} {\bf 101}, 245156 (2020).

\bibitem{clementyev2009} E.~S. Clementyev, and A.~V. Mirmelstein, {\it Electronic Properties of Solids} {\bf 136}, 128-139 (2009).








\bibitem{binder2980} K. Binder, Ordering of the face-centered-cubic lattice with nearest-neighbor interaction. {\it Phys. Rev. Lett.} {\bf 45}, 811-814 (1980).

\bibitem{moessner2003} R. Moessner, S. L. Sondhi, Three-dimensional resonating-valence-bond liquids and their excitations. {\it Phys. Rev. B} {\bf 68}, 184512 (2003).









\bibitem{footnote}
Another factor that could potentially 
contribute to the robustness of the pseudogap state against magnetic 
fields is the small Land\'{e} {\it g}-factor~\cite{jensen1991} of $g=\frac{2}{7}$ 
expected for a $5f$-electron shell consisting of five electrons~\cite{zhu2007}.

\bibitem{jensen1991} J. Jensen, A. R. Mackintosh, The international Series of Monographs on Physics: Rare Earth Magnetism; Structures and Excitations {\it eds. J. Borman, S.. F. Edwards, C. H. Llewwellyn Smith, M. Rees}  (Clarendon Press, Oxford, 1991).

\bibitem{sarrao2002} J. L. Sarrao, L. A. Morales, J. D. Thompson, B. L. Scott,
G. R. Stewart, F. Wastin, J. Rebizant, P. Boulet, E. Colineau, G. H. Lander, Plutonium-based superconductivity with a transition temperature above 18K. {\it Nature} {\bf 420}, 297-299 (2002).

\end{thebibliography}

\end{document}